# Privacy Knowledge Modelling for Internet of Things: A Look Back

*Charith Perera (Open University), Chang Liu (CSIRO), Rajiv Ranjan (Newcastle University), Lizhe Wang (China University of Geosciences), Albert Y. Zomaya (University of Sydney)*

## Abstract

Internet of Things (IoT) and cloud computing together give us the ability to sense, collect, process, and analyse data so we can use them to better understand behaviours, habits, preferences and life patterns of users and lead them to consume resources more efficiently. In such knowledge discovery activities, privacy becomes a significant challenge due to the extremely personal nature of the knowledge that can be derived from the data and the potential risks involved. Therefore, understanding the privacy expectations and preferences of stakeholders is an important task in the IoT domain. In this paper, we review how privacy knowledge has been modelled and used in the past in different domains. Our goal is not only to analyse, compare and consolidate past research work but also to appreciate their findings and discuss their applicability towards the IoT. Finally, we discuss major research challenges and opportunities.

## Introduction

The Internet of Things allows people and things to be connected anytime, anyplace, with anything and anyone, ideally using any path, any network and any service [1]. Over the last few years, a large number of Internet of Things (IoT) solutions have come to the IoT marketplace [2]. These IoT solutions together collect a large amount of data that can be used to derive useful but extremely personal knowledge about users [3]. On the other hand, cloud computing provides ubiquitous, convenient, on-demand access to a shared pool of configurable computing resources. As such services are becoming more powerful and cheaper, the risk of users' privacy being violated due to knowledge discovery increases significantly.

A business model that has been developed to derive value out of such data is *open data markets* [3]. The idea of an *open data market* envisions the creation of a data trading model to facilitate exchange of data between different parties in the Internet of Things (IoT) domain. The data collected by IoT products and solutions are expected to be traded in these markets. Data owners will collect data using IoT products and solutions. Data consumers who are interested will negotiate with the data owners to get access to such data. Data captured by IoT products will allow data consumers to further understand the preferences and behaviours of data owners and to generate additional business value using different techniques ranging from waste reduction to personalized service offerings. In open data markets, data consumers will be able to give back part of the additional value generated to the data owners. However, privacy becomes a significant issue when data that can be used to derive extremely personal information is being traded. Therefore, it is important to understand what privacy is and what privacy means for each user of a given system in order to design the systems to ensure privacy is protected at all times.

One of the widely accepted definitions, presented by Alan F. Westin [4], describes information privacy as "the claim of individuals, groups or institutions to determine for themselves when, how, and to what extent information about them is communicated to others". Knowledge modelling is a process of creating a computer interpretable model of knowledge or standard specifications about a process, a product, or a concept. In this paper, our focus is on *'privacy'*. We consider any piece of information that can be used to understand the privacy expectation of an individual in any given context as privacy knowledge.

Our objective in this paper is to survey how privacy knowledge has been modelled in the past in different domains. It is important to note that we do not intend to review an exhaustive amount of past work, but to capture insights from a broad range of approaches. We also discuss how past approaches can be used or are relevant in the IoT domain. In web domain, only the users' online activities are captured. In contrast, IoT systems can capture users' activities and behaviours 24/7 through various kinds of devices. Therefore, IoT domain poses significant privacy risks compared to web domain. We also analyse different privacy modelling approaches to identify any common patterns and applications.

In the rest of this paper, we briefly look back at major attempts in the past towards giving privacy control to users. Next, we examine how privacy knowledge has been modelled by researchers in the past including the factors they have considered in their models, techniques used to implement it, application domains and so on. Then, we present lessons learnt from our review by identifying major highlights from past work and providing insights into future work. Finally, we discuss a few major research challenges, namely the importance of developing a comprehensive privacy knowledge model for IoT and the importance of developing non-intrusive user privacy preferences knowledge acquisition techniques, before we conclude the paper.

## Privacy Knowledge Modelling: Historical View

In this section, we review one of the major privacy preference modelling approaches of the past, namely Platform for Privacy Preferences (P3P) (w3.org/P3P) [5]. We consider P3P as a key milestone of modelling privacy preference despite its failure due to various reasons as discussed later. P3P is not designed for IoT but for the web domain where it can only attempt to protect user privacy during online web browsing activities. Further, it is important to understand what P3P is, how it has been designed to work, and why P3P failed in order to propose the next generation privacy preferences modelling approaches, especially for newer paradigms such as open data markets in IoT. Examination of P3P will help us to understand the challenges in privacy modelling and eventually privacy management.

P3P is a protocol allowing websites to declare their intended use of information they collect about web browser users. P3P is not an ontology-based model, but an XML mark-up language based description model. The initial intention was to give users more control of their personal information when web browsing. P3P is a machine-readable language that helps to express a website's data management practices. In P3P, information is managed based on privacy policies of the users as well as the websites.

P3P works as follows. First, the websites specify a set of policies defines their intended use of personal information that may be gathered from their site visitors. From the other end, users are required to define their own set of preferences for collection and processing of personal information by the sites they visit. When a user visits a site, P3P compares the user's policy with the policy of the website. The comparison primarily evaluates what personal information the user is willing to release, and what information the website wants to receive. However, if the two sets of privacy policies do not match, P3P will advise users and ask if they are willing to proceed to the site despite the risk of giving up more personal information. It is important to note that users are in a somewhat helpless situation here with limited options to proceed. P3P is designed as a way to express privacy preferences but not as a negotiation framework. Therefore, P3P privacy profiles are designed to be static. For example, a user may define a privacy policy saying that information about their browsing habits should not be collected. If the policy of a website states that a cookie is used for this purpose, the browser automatically rejects the cookie. However, it is likely that key parts of the website's functionality will depend on this cookie and by rejecting it the user's experience will be degraded. As a result, most of the time, websites tend to get the information they want. The only exception would be if users in large numbers decide not to visit a particular website due to that website's unreasonable privacy policies; the website may implicitly be pressured to change its own policies. Figure 1 summarizes the main content of a privacy policy in P3P.

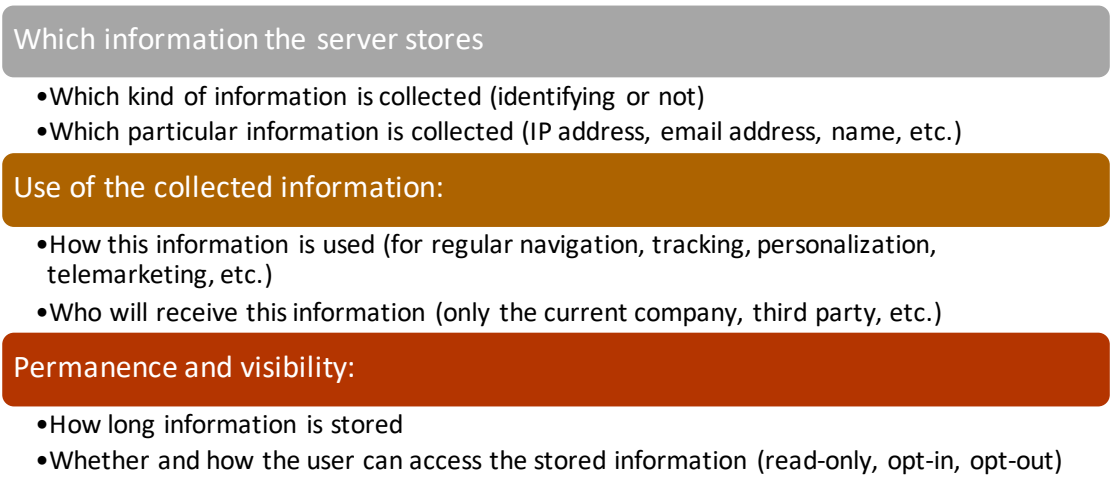

Figure 1: Main Content of P3P

Despite its lack of adoption due to various technical difficulties and lack of value, P3P brings an important ideology of allowing users to define their privacy expectations. P3P puts users in control of their privacy. Ideologically, P3P and our proposed knowledge model share objectives for expressing privacy. Machine interpretability is one of the key objectives in P3P. For example, once privacy preferences are defined by both websites and users, remaining interactions could occur almost autonomously with minimum human intervention. This is also one of the main priorities in open data markets as we envision a large number of data trading transactions.

In open data markets, data owners and consumers should be able to collectively define their privacy preferences and policies in such a way that machines can take over the trading negotiations and act on behalf of data owners and consumers. Building a common knowledge model using ontologies increases the ability to conduct trading activities autonomously. P3P intended to create an open and transparent method to express privacy preferences. It also intended to make it easier for users to set privacy policies.

Despite all the potential benefits, P3P failed [6] to receive the necessary attention from browser makers, Internet advocates and institutions. The Electronic Privacy Information Center (EPIC) identifies P3P as a "*complex and confusing protocol that will make it more difficult for Internet users to protect their privacy*" [6]. One of the drawbacks in P3P is that it will effectively exclude good web sites that lack P3P code even though the privacy practices of these sites may far exceed sites that are P3P-compliant. Another challenge is the lack of any means to enforce privacy policies. Another criticism towards P3P is over a lack of effective ways of educating people on the level of privacy and what P3P actually does to protect people. As mentioned earlier, this make it harder for non-technical users to understand and configure P3P based on privacy expectations.

Now let us discuss why we need something similar to P3P in open data markets in the IoT domain in order for data trading to occur and why such an approach would work despite P3P being a failure. It is also important to identify ways to overcome the above mentioned issues that disrupted the adoption of P3P. In open data market environments, each data owner may have his/her own privacy preferences. The data they would like to trade with other parties may vary based on number of factors such as type of data, type of data consumer, the purpose of data collection, how the data is managed, risks involved, expected return, and so on. From data consumers' point of view, factors such as accuracy of the data, frequency of the data capture, frequency of the data communication, the amount of value that can be generated by using the data, play a significant role when deciding whether to buy data from a certain data owner.

As mentioned earlier, data consumers are not interested in buying data from one or two data owners. In order to derive useful knowledge, data consumers need to gather process data on a large scale (i.e. >10,000). Such knowledge will help data consumers to save substantial amounts of waste or to generate new customer value. Part of such value will be given back to data owners to attract them again as potential data sellers. In traditional market settings, a trade would occur when the buyer perceives a certain product (or a service) as equally or more valuable to him in comparison to the price the seller is expecting in return. Similarly, a data trade between a data owner and a data consumer depends on perceived privacy risks and benefits. If the data owners perceive that they get a return sufficient to trade off with the privacy risks involved, they will agree to sell their data.

In open data markets, we intend to use privacy preferences models to conduct data trading negotiations between data owners and consumers instead of making strict decisions based on static privacy profiles. That means both data owners and consumers may be willing to change their privacy expectations based on privacy risks and rewards involved in each trading occurrence. This is a different approach compared to the P3P approach where static privacy preferences are compared between websites and users. In order to conduct data trading as well as to perform automated risk benefits negotiations, we need to capture and model a certain amount of information from both data owners and consumers.

To support P3P, W3C has also developed and recommended a language called APPEL (A Privacy Preference Exchange Language) (w3.org/TR/P3P-preferences) that can be used to express user privacy preferences. Using the APPEL language, users are allowed to express their privacy preferences using a set of preference-rules, which can then be used by software to make automated or semi-automated decisions regarding the acceptability of machine-readable privacy policies from P3P enabled Web sites. The major weakness of APPEL is that it can specify only what is *unacceptable* for a user, but not what is *acceptable* for a user. The IoT demands a more comprehensive approach towards privacy modelling, purely because it envisions to deal, not only with our online lives, but also with offline everyday lives. Due to this, privacy risks will grow exponentially.

## Evaluation of Research Efforts: Modelling and Applications

In the previous section, we discussed why modelling privacy is important, how privacy has been handled in the past, and identified the major challenges in modelling and managing user privacy in a web browsing context. In this section, we review a broad range of past experiences in modelling privacy knowledge in different application contexts.

Non-Ontology based Privacy Knowledge Modelling

- Jaroucheh at al. [7] proposed a context information dissemination framework based on privacy policies. Their application domain is a smart university where university staff and students across different universities who collaborate with each other get to keep updated about each other's current activities, status and interests and to exchange information so that they can avoid disturbing and interact more efficiently. Context information about each person is gathered through online services such as Google Wave. Custom defined XML schemas have been used to model user privacy requirements and the users in the system get to decide which consumer is allowed to access their context information (e.g. location) at any given time. Context information is protected by these policies and only released to the authorised personnel.

Ontology based Privacy Knowledge Modelling

- Zhang and Todd [8] have developed a privacy ontology for context-aware systems. Towards developing this *privacy ontology*, they have adopted P3P terminology and created corresponding classes and properties. A *Privacy Rule* class is defined to represent privacy preferences set by users. Every privacy rule is expressed with two elements: *Data* (Data class) and *Conditions* (Condition class). The *Conditions* class contains all conditions under which a user is willing to disclose data. As also mentioned in the P3P specification, the conditions can be classified based on various individual preferences including recipients of data, purposes of data collection, duration that data will be kept by recipients, a user's access privilege to his personal data once stored by recipients, and ways of handling disputes. In this approach each data item (or collection) is attached to a rule that comprises of set of conditions as mentioned above.

- Sacco and Passant [9] have proposed a lightweight vocabulary built on top of Web Access Control (WAC) (w3.org/wiki/WebAccessControl) that enables users to create fine-grained privacy control for their data. Similar to the above approaches, the access restrictions are put in place on an individual resources level (i.e. document level not data items within the document level). Web Access Control is a vocabulary that defines access control privileges in web documents.

- Hu and Yang [10] have also adopted a similar ontology where resources are protected at resource level through conditions. They have modelled factors such as purposes (allowed and not allowed) to use data, use time period for data retention, who (i.e. data consumers) allowed to access data and not allowed to access data, and obligations. One of the highlights in this work is that capturing is both allowed and not allowed separately (e.g. users and purposes). This provides an additional opportunity when evaluating a given data request and certain information is not evaluable. As a result, global rules can be defined to allow a data request if a data owner has not explicitly defined certain factors. This may also combine with other factors. An example condition would be if the data consumer is a research institute, allow data even if they have not defined the data retention period exactly.

- Kost et al. [11] have gone a step beyond to identify ten factors that need to be captured by privacy preferences modelling namely, purpose, consent, limited collection, limited use, disclosure, retention, accuracy and context preservation, security, openness, and compliance. In comparison to above approaches, this work has highlighted the importance of capturing a much broader range of information about a particular data collection and analysis task.

- Ahmed et al. [12] included a much broader set of privacy concepts in their knowledge model. A highlight of this work is modelling knowledge that describes how data will be treated during the data communication, transfer, storage and data processing. PROACT [13] is an ontology designed for ambient environments which also captures detailed information about data processing mechanisms such as data collection, data transfer, data storage and so on.

- In their work, Youssef et al. [14] proposed a model to capture privacy preferences of mobile consumers. One of the significant features of this ontology is that it captures incentives. They analysed different types of promotion and found the following classes: 1) monetary, 2) coupon, 3) time slack, 4) extra items, and 5) payment on instalments. In relation to each incentive class, users get to specify expected values as well.

- Bodorik et al. [15] have also proposed a privacy preferences model. It allows users to place restriction based on factors such as purpose of data usage, data recipient (i.e. data consumer), data retention, sdispute, remedy, and access control (i.e. who has access to data). Some of the purposes they have listed are admin, develop, tailoring, pseudo-analysis, pseudo-decision, individual-analysis, individual-decision, contact, historical, telemarketing and other-purpose. Similarly, some of the options under retention is no-retention, stated-purpose, legal-requirement, indefinitely, business-practices. Similar retention details are presented by [16] and [17] as well.

- Rei [18] is a highly expressive policy language which enables users to specify their privacy preferences. *Rule Priority* is an interesting concept presented in Rei which enables the ability to define different combinations of conditions with different outcomes. In contrast, previous approaches only allowed one rule with multiple conditions where all the conditions need to be met in order to get access to data. Such expressions are very important in data market negotiations where users may want to define different sets of privacy preferences conditions based on the type of the data consumer. For example, a data owner may expect the data consumer to perform limited knowledge discovery if the data consumer is a commercial entity. However, the same data owner may grant unlimited knowledge discovery for not-for-profit medical research institutes based on his personal views, beliefs, and value systems.

- To a growing list of factors to be modelled in an *ideal* privacy knowledge model, Martimiano et al. [19] proposed the inclusion of trust levels. Even though the paper vaguely explained what trust levels are with fixed sets of classes with static assignments such as close family, unknown person, known person, close friend, co-worker. However, this scheme of categorising is most suited to the Facebook type categorization where an individual can only be a friend or not a friend. In contrast Martimiano et al.'s approach is much more aligned with how social interactions actually work. Another work that has highlighted the importance of modelling trust is [20].

So far we evaluated a number of different privacy knowledge modelling approaches. In Table 1, we summarize all of these discussions by listing each approach, their primary application domain, and the factors they have identified as important for inclusion in their knowledge models.

*Table 1: A Summary of factors modelled in past privacy knowledge modelling approaches*

| Citation | Modelling Language Used | Primary Application Domain | Data | Data Consumer | Time / Retention Period | Purpose | Dispute / Remedy | Access | Policy / Disclosure | Actions | Obligations / Limits | Data Consumer(Opt | Purpose (Opt out) | Consent | Techniques Used* | Incentive | Trust |
|---|---|---|---|---|---|---|---|---|---|---|---|---|---|---|---|---|---|
| [7] | XML | Smart University | x | x | x | | | | | | | | | | | | |
| [8] | Ontology | Ubiquitous Computing | x | x | | x | x | x | | | | | | | | | |
| [9] | Ontology | Linked Data | x | x | | x | | x | | | | | | | | | |
| [10] | Ontology | Healthcare | x | x | x | x | | x | x | x | x | x | x | | | | |
| [11] | Ontology | Transportation | x | x | x | x | | x | x | | x | | | x | | | |
| [12] | Ontology | Personal Information System | x | x | x | x | | | x | | | | | | CTS | | |
| [13] | Ontology | Mobile E-commerce | x | x | x | | | | | | | | | | | x | |
| [14] | Ontology | - | x | x | x | x | | x | | | | | | | | | |
| [15] | Ontology | Web Services | x | x | x | x | x | x | | | x | x | | | | | |
| [16] | Ontology | E-commerce | x | x | x | x | | | x | | | | | x | | | |
| [18] | Ontology | Ubiquitous Computing | x | x | | x | | | x | | | | | | | | x |
| [17] | Ontology | Service Oriented Architecture | x | x | x | x | | x | x | x | x | | | x | C | | |
| [19] | Ontology | Ubiquitous Computing | x | x | x | x | | x | x | x | | | | | CTS | | x |

*Techniques used: Data Collection (C), Data Transmission (T), Data Storage (S)

**Lessons Learnt**

One of the important trends we observe is the increasing adoption of ontologies (e.g. OWL, RDF) to model privacy knowledge [8]. This trend is especially true with recent work. Other than ontologies, researchers have also used custom defied XML schemes [7] to mode privacy. For example, the P3P approach discussed earlier is driven by an XML schema. An ontology defines a common vocabulary for researchers who need to share information in a domain. They allow capture of the meaning between different concepts. It includes machine-interpretable definitions of basic concepts in the domain and relations among them. Further, ontologies promote the reuse of domain knowledge. In contrast, XML-like markup languages are typically used to structure data but not able to capture the semantics and relationships. As discussed earlier, one of the main weaknesses in the P3P approach was difficulty in arriving at an agreed-upon vocabulary. Ontologies can address this issue reasonably well as it allows modeling of relationships between similar terms so the agreements are not critical compared to markup language based modeling.

Another advantage of employing ontologies to model privacy knowledge is that they allow privacy policies to be defined at both data level (i.e. instance level) and class level which is very convenient for the users. Instance level rules should be given priority and class level rules could be used in the absence of instance level rules.

Privacy preferences of users may change over time where ideal systems should be able to adapt autonomously. Especially in IoT as well as in open data market scenarios, it is important to understand users' privacy needs proactively and predict their preference ahead of time so the data owners do not have to deal with privacy configurations.

One of the common weaknesses in these approaches is the lack of support to capture and model information about data management techniques. For example, what techniques are used to store data (e.g. encryptions techniques), how data will be routed (e.g. torproject.org), are important factors for data owners as they may have direct impact on their opinion on whether to share data with a particular data consumer or not. Even though some approaches have highlighted the importance of modelling data management related information, proper technique-level vocabulary and concepts are not being introduced.

In general, most of the approaches use some kind of privacy policy in some form such as rules, preferences, conditions, and so on. Policies typically define who can access a certain resource and under which conditions, how data should be provided to data consumers, and how the provided information will be used and so on. Such privacy preference configuration could be exhaustive, if done at data item level. However, usage of ontologies makes this somewhat simpler by supporting class level policy definitions

Based on a number of privacy regulations such as US, EU and OECD, Garcia et al. [16] have presented a number of privacy requirements that need to be considered when developing a privacy knowledge model. In Figure 2, we expanded their requirement list based on the lessons learned by evaluating past approaches as presented in Table 1. We also recommend to support the needs of IoT and open data markets.

## Future Research Directions, Challenges and Opportunities

In this section, we briefly highlight a few major research challenges that need to be addressed in the future, with a particular emphasis on the needs of the IoT [1] and open data markets [3].

**Privacy Preferences Modelling and User Profiling**
Privacy Preferences profiling is the task of modelling user preferences in a common structure. There are many factors that could impact a user's privacy preferences especially in the IoT domain as well as in open data market scenarios. One major challenge is to find all kinds of factors that could make an impact on users' minds when they think about their privacy expectations. For example, when participating in open data markets [3], some users (i.e. data owners) may consider the reputation of data consumers as the most important factor to be considered when making their decisions whether to trade data or not. However, for some users, the purpose of the data collection may be the most significant factor. At the same time, some factors could be completely meaningless to some users depending on their level of technical knowledge. For example, the type of encryption supported by a

- A description of what data are collected and how they are used by data consumers should be available to data owners.
- Data owners should be able to agree with the collection of their data before it happens.
- The techniques used to collect a data item should be identified.
- The collector (e.g. Data Consumer) of a data item should be identified.
- The purposes for which a data item is collected should be identified.
- The entities (e.g. third party) to which a data item is disclosed by its collector should be identified.
- The data items to be collected should be identified.
- The retention time of a data item should be identified.
- Data consumers should indicate if data owners are allowed to complete, correct and update their retained data.
- Data consumers should indicate if data owners can request records on how their data have been used, in formats understandable by data owners and with known delays and charges.
- Data consumers should indicate if data owners are able to request copies of data on them, in formats understandable by data owners and with known delays and charges.
- Data consumers should clearly inform data owners regarding what kind of knowledge is expected to be discovered using their data.
- Data owners should know the risks, their impact level of sharing (trading) a particular type of data before sharing (trading) occurs.
- Data owners and the data consumers should come to an agreeent regarding the reward that the data owners may receive as a return for taking the risks of sharing their data.
- Reward types associated with sharing data need to be identified clearly before any data sharing would occur.
- Data owners should be able to apply *data quality reduction* techniques before data is being sent to the data consumers to reduce privacy risks.
- Both data owners and data consumers should agree on which *data quality reduction* techniques will be used.

*Figure 2: Privacy Preferences Modelling Requirements [The proposed extensions are marked in green]*

given data consumer may have no impact on the mindset of non-technical users as they are not capable of evaluating and understanding the value of encryption.

The advantage of modelling privacy knowledge of each user is that it allows both humans and machines to share a common vocabulary. Modelling the privacy preferences in a machine interpretable way enable development of smart systems that would automatically understand user preferences and act accordingly. For example, in highly dynamic environments such as the IoT, automated configuration of privacy preferences would be significantly helpful for users as it might reduce the users' workload as well as privacy concerns. Further, a common understanding will help different software programs to use the privacy knowledge model to provide different types of value added services such as proactive privacy preferences configuration, learn user behaviour over time and predict users' privacy expectations and so on. Additionally, a common privacy preference knowledge model would be helpful in conducting data trading negotiations in open data market environments [3]. In summary, an ideal privacy knowledge model should be able to capture any piece of information that would potentially make an impact on privacy. Secondly, such models should be able to capture users' priorities where each user may treat different factors with different priorities.

**User Privacy Preferences Acquisition**

In addition to building privacy preferences knowledge models, it is also important to develop techniques that can be used to acquire preferences from users in a non-intrusive manner. Once we have the privacy preferences knowledge model, which can be considered as the template, the next challenge is to acquire users' (e.g. data owners') privacy preferences with minimum human intervention. Asking for too much information about preferences from data owners may overload them, while a lack of information could lead to a violation of their privacy expectations. Recommender systems could be useful in addressing this issue where a basic template for each user may be built by analyzing and studying similar users (e.g. demographics). Towards this, limited questions and answer mechanism can be employed to identify users' personalities so recommender systems can predict some parts of the privacy preferences. Then, users can be questioned again to fill the remaining essential privacy preferences parameters. One of the main challenges from data consumers' perspective is scalability and cost associated with data acquisition. Data acquisition negotiations need to be done individually with each data owner. Even though a single data trading transaction may not consume substantial amount of computational resources, large numbers of such transactions will surely do. Therefore, data acquisition negotiations algorithms should be efficient and scalable.

## Conclusions

This survey provides an overview of how privacy knowledge of users has been modelled in different domains in the past. We presented different techniques of privacy preference modelling techniques from XML based mark-up languages to more semantically enriched ontologies. We also reviewed different privacy related concepts captured and modelled by different projects to support their unique needs. After comparing projects, we identified some common weaknesses in the existing approaches in modelling privacy knowledge and why the Internet of Things domain demands a more comprehensive privacy knowledge modelling approach. Finally, we discussed major research challenges and opportunities. We recommend to develop a comprehensive knowledge model that is capable of capturing user privacy knowledge. Specially, we argue that ontologies represent the most appropriate method of modelling privacy knowledge due to its ability to support modelling of relationships between concepts and automated reasoning. Further, we recommend conducting research into developing techniques that are capable of acquiring privacy preferences autonomously, with limited intervention from users, to avoid overloading them.

## Acknowledgement

Dr. Charith Perera's work is funded by European Research Council Advanced Grant 291652 (ASAP).

# References

[1] C. Perera, A. Zaslavsky, P. Christen and D. Georgakopoulos, "Context Aware Computing for The Internet of Things: A Survey," *Communications Surveys Tutorials, IEEE,* vol. 16, no. 1, pp. 414-454, 2013.

[2] C. Perera, C. Liu and S. Jayawardena, "The Emerging Internet of Things Marketplace From an Industrial Perspective: A Survey," *Emerging Topics in Computing, IEEE Transactions on,* vol. 3, no. 4, pp. 585-598, 2015.

[3] C. Perera, R. Ranjan and L. Wang, "End-to-End Privacy for Open Big Data Markets," *Cloud Computing, IEEE,* vol. 2, no. 4, pp. 44-53, July 2015.

[4] A. F. Westin, Privacy and Freedom, New York: The Bodley Head Ltd, 1967.

[5] L. F. Cranor, A. M. McDonald, S. Egelman and S. Sheng, "2006 Privacy Policy Trends Report," CyLab Carnegie Mellon University, Pittsburgh,, 2007.

[6] Electronic Privacy Information Center, "Pretty Poor Privacy: An Assessment of P3P and Internet Privacy," epic.org, 2000. [Online]. Available: https://epic.org/reports/prettypoorprivacy.html. [Accessed 01 10 2015].

[7] Z. Jaroucheh, X. Liu and S. Smith, "An Approach to Domain-based Scalable Context Management Architecture in Pervasive Environments," *Personal Ubiquitous Comput.,* vol. 16, no. 6, pp. 741-755, #aug# 2012.

[8] N. Zhang and C. Todd, *Developing a privacy ontology for privacy control in context-aware systems,* Dept. of Electronic & Electrical Engineering, University College London, 2006.

[9] O. a. P. A. Sacco, "A Privacy Preference Ontology {(PPO)} for Linked Data," in *Workshop on Linked Data on the Web, Hyderabad, India, March*, 2011.

[10] Y.-J. Hu and J.-J. Yang, "A Semantic Privacy-preserving Model for Data Sharing and Integration," in *Proceedings of the International Conference on Web Intelligence, Mining and Semantics*, New York, NY, USA, 2011.

[11] M. Kost and J. C. Freytag, "Privacy Analysis Using Ontologies," in *Proceedings of the Second ACM Conference on Data and Application Security and Privacy*, New York, NY, USA, 2012.

[12] M. Ahmed, A. Anjomshoaa and A. M. Tjoa, "Context-based Privacy Management of Personal Information Using Semantic Desktop: SemanticLIFE Case Study," in *Proceedings of the 10th International Conference on Information Integration and Web-based Applications \& Services*, New York, NY, USA, 2008.

[13] I. Panagiotopoulos, L. Seremeti, A. Kameas and V. Zorkadis, "PROACT: An Ontology-Based Model of Privacy Policies in Ambient Intelligence Environments," in *Informatics (PCI), 2010 14th Panhellenic Conference on*, 2010.

[14] M. Youssef, N. R. Adam and V. Atluri, "Semantically Enhanced Enforcement of Mobile Consumer's Privacy Preferences," in *Proceedings of the 2006 ACM Symposium on Applied Computing*, New York, NY, USA, 2006.

[15] P. Bodorik, D. Jutla and M. X. Wang, "Consistent Privacy Preferences (CPP): Model, Semantics, and Properties," in *Proceedings of the 2008 ACM Symposium on Applied Computing*, New York, NY, USA, 2008.

[16] D. Garcia, M. Toledo, M. Capretz, D. Allison, G. Blair, P. Grace and C. Flores, "Towards a base ontology for privacy protection in service-oriented architecture," in *Service-Oriented Computing and Applications (SOCA), 2009 IEEE International Conference on*, 2009.

[17] M. Hecker, T. S. Dillon and E. Chang, "Privacy Ontology Support for E-Commerce," *Internet Computing, IEEE,* vol. 12, no. 2, pp. 54-61, March 2008.

[18] L. Kagal, T. Finin, M. Paolucci, N. Srinivasan, K. Sycara and G. Denker, "Authorization and privacy for semantic Web services," *Intelligent Systems, IEEE,* vol. 19, no. 4, pp. 50-56, Jul 2004.

[19] L. Martimiano, M. Goncalves and E. dos Santos Moreira, "An ontology for privacy policy management in ubiquitous environments," in *Network Operations and Management Symposium, 2008. NOMS 2008. IEEE*, 2008.

[20] Z. Iqbal, J. Noll, S. Alam and M. Chowdhury, "Toward User-Centric Privacy-Aware User Profile Ontology for Future Services," in *Communication Theory, Reliability, and Quality of Service (CTRQ), 2010 Third International Conference on*, 2010.

**Short Author Bios**

**Charith Perera** is a Research Associate at The Open University, UK. Currently, he is working on the Adaptive Security and Privacy (ASAP) research programme. He received his BSc (Hons) in Computer Science in 2009 from Staffordshire University, Stoke-on-Trent, United Kingdom and MBA in Business Administration in 2012 from University of Wales, Cardiff, United Kingdom and PhD in Computer Science at The Australian National University, Canberra, Australia. Previously, he worked at Information Engineering Laboratory, ICT Centre, CSIRO. His research interests are Internet of Things, Sensing as a Service, Privacy, Middleware Platforms, Sensing Infrastructure. He is a member of both IEEE and ACM. Contact him at charith.perera@ieee.org

**Chang Liu** is currently a postdoctoral researcher associate in the Data61 Group at CSIRO, Australia. He received his PhD degree in computer science in 2015 from University of Technology, Sydney, Australia. He received his B.Eng. degree in computer science in 2005 and M.Sc. degree in information security in 2008, both from Shandong University, Jinan, China. His research interests include cloud computing, big data, data security, data privacy, and applied cryptography. He has published his research in many top level venues such as IEEE Transactions on Parallel Distributed Systems and IEEE Transactions on Computers. Contact him at chang.liu@csiro.au

**Rajiv Ranjan** is an associate professor in the School of Computing Science at Newcastle University. He is an internationally established scientist with about 160 publications and expertise in cloud computing, big data, and Internet of Things. Before moving to Newcastle University, he was Julius Fellow, Senior Research Scientist and Project Leader in the Digital Productivity and Services Flagship of Commonwealth Scientific and Industrial Research Organization. Ranjan received a PhD in computer science from the University of Melbourne. He serves on the editorial boards of top quality international journals including IEEE Transactions on Computers, IEEE Transactions on Cloud Computing, IEEE Cloud Computing, and Future Generation Computer Systems. He is a member of IEEE. Contact him at raj.ranjan@ncl.ac.uk.

**Lizhe Wang** is a "ChuTian" Chair Professor at School of Computer Science, China University of Geosciences (CUG), and a Professor at Institute of Remote Sensing & Digital Earth, Chinese Academy of Sciences (CAS). Prof. Wang received B.E. and M.E from Tsinghua University and Doctor of Engineering from University Karlsruhe (Magna Cum Laude), Germany. Prof. Wang is a Fellow of IET, Fellow of British Computer Society. Prof. Wang serves as an Associate Editor of IEEE T. Computers, IEEE T. on Cloud Computing, IEEE T. on Sustainable Computing. His main research interests include HPC, e-Science, and remote sensing image processing. Contact him at lzwang@ceode.ac.cn.

**Albert Y. Zomaya** is a Chair Professor and Director of the Centre for Distributed and High Performance Computing at Sydney University. Professor Zomaya published more than 500 scientific papers and is author, co-author or editor of more than 20 books. He is the Editor in Chief of the IEEE Transactions on Sustainable Computing and serves as an associate editor for 22 leading journals. Professor Zomaya is the recipient of the IEEE TCPP Outstanding Service Award (2011), the IEEE TCSC Medal for Excellence in Scalable Computing (2011), and the IEEE Computer Society Technical Achievement Award (2014). He is a Fellow of AAAS, IEEE and IET. Contact him at albert.zomaya@sydney.edu.au.